\documentstyle[manuscript,aps,12pt,epsfig]{revtex}

\newcommand{\bm}{\bibitem}

\begin{document}

\draft
\tighten

\title{ Role of $N^*$(1650) in the near threshold $pp \to p\Lambda K^+$
and $pp \to p\Sigma^0 K^+$ reactions\footnote{Work supported by BMBF and GSI,
Darmstadt} } 
\date{\today}
\author{R. Shyam$^{a,c}$, G. Penner$^b$, and U. Mosel$^b$}
\address {$^a$ 
Saha Institute of Nuclear Physics, Calcutta 700064,
India \\
$^b$ Institut f\"ur Theoretische Physik, Universit\"at Giessen,
D-35392 Giessen, Germany \\
$^c$ National Superconducting Cyclotron Laboratory, Michigan State University,
East Lansing, MI 48814, U.S.A.} 
\maketitle

\begin{abstract}
We investigate the $pp \to p\Lambda K^+$ and $pp \to p\Sigma^0 K^+$
reactions at beam energies near their thresholds within
an effective Lagrangian model, where 
the strangeness production proceeds via the excitation of 
$N^*$(1650), $N^*$(1710), and $N^*$(1720) baryonic resonances. It is  
found that the $N^*$(1650) resonance dominates both these reactions 
at near threshold energies. The contributions from this resonance
together with the final state interaction among the outgoing particles
are able to explain the observed beam energy dependence of the ratio
of the cross sections of the two reactions in the near threshold region. 
\end{abstract}
\newpage
Recently, at the Cooler Synchrotron (COSY) facility in J\"ulich 
measurements have been performed \cite{bal98,sew99} for the
associated strangeness production
in proton-proton ($pp$) collisions at near threshold beam energies.
A very interesting result of these studies is that the ratio (R) of the 
total cross sections for the $pp \to p\Lambda K^+$ and
$pp \to p\Sigma^0 K^+$ reactions (to be referred as $\Lambda K^+$
and $\Sigma^0 K^+$ reactions respectively) at the same excess energy
(defined as $\epsilon = \sqrt{s}-m_p-m_Y-m_K$, with $m_p$, $m_Y$, and
$m_K$ being the masses of proton, hyperon, and kaon respectively and $s$
the invariant mass of the collision), is about $28^{+6}_{-9}$ for
$\epsilon$ $<$ 13 MeV. This result is very intriguing because at higher beam
energies \cite{lan88} this ratio is only around 2.5.

Assuming that the hyperon production
proceeds solely due to the kaon($K$)-exchange mechanism and that the
final state interaction (FSI) effects among the outgoing particles are
absent, R is given essentially by the ratio
($g^2_{\Lambda N K}/g^2_{\Sigma N K}$) of the squares of coupling constants
at the vertices from which the $K^+$ meson emerges. Although, values of 
$g_{\Lambda N K}$ and $g_{\Sigma N K}$ are not known with certainty
\cite{lik98}, yet the SU(6) prediction of this ratio \cite{swa63} is
27 which would nearly explain the observed value of R. However,
$\pi$-exchange mechanism is shown \cite{lik98,shy99,lag91,sib97,kai99}
to be important for these reactions. The two mechanisms taken  
together lead to a considerably lower value \cite{sew99}($\sim$ 3.6)
for R. Another qualitative explanation \cite{sew99} of these data 
suggests that the dominant $\Sigma-p$ final state interaction, which
includes the
$\Sigma N \to \Lambda N$ conversion process, suppresses the $\Sigma^0$
production. Although some support in favor of this conversion does exist 
\cite{cli68}, it is not evident that the whole
of the observed enhancement is really due to the produced $\Sigma$
particle being converted to $\Lambda$ by the FSI effects. 

Recently, a few quantitative calculations have been reported to
explain this result. 
Assuming that the $\pi$- and $K$- exchange processes are the only mechanism
leading to the strangeness production, the authors of Ref. \cite{gas00}
show within a (non-relativistic) distorted wave Born approximation (DWBA)
model that while the $\Lambda K^+$ reaction is
dominated by the $K$-exchange only, both $K$- and $\pi$-
exchange processes play an important role in the case of 
$\Sigma^0 K^+$ reaction. Therefore, if the amplitudes
corresponding to the two exchanges in the latter case interfere
destructively, the production of $\Sigma^0$ is suppressed as compared
to that of $\Lambda$. It is also shown in Ref. \cite{gas00} that
FSI effects, although important, can not explain the large value of R
on their own. However, a conclusive evidence in support of the  
relative signs of $\pi$- and $K$- exchange amplitudes being
opposite to each other is still lacking. Furthermore,
other mechanisms like excitation, propagation, and decay of intermediate
baryonic resonances play (see eg. \cite{shy99,col97}) an important role in
the strangeness production, which may change the scenario of Ref. \cite{gas00}.
It is also not clear if this model can simultaneously explain the relatively
smaller value of R at larger beam energies (i.e. for $\epsilon$ $\sim$ 1 GeV).

In Ref. \cite{sib00} two types of boson exchange models have been used to
calculate the ratio R. In one of them, the strangeness production 
proceeds solely via $\pi$- and $K$- exchange mechanisms. Neglecting the
interference between the corresponding amplitudes 
and making corrections for the FSI effects via the Jost
function method of the Watson-Migdal theory \cite{wat52}, the predictions
of this model are found to be in agreement with the observed ratio within
a factor of 2 in the near threshold region.
However, these authors find $K$- and $\pi$- exchange amplitudes
to be of similar magnitudes for both $\Lambda K^+$ and $\Sigma^0K^+$
reactions in the near threshold region, which in disagreement
with the results of Ref. \cite{gas00}.
Moreover, form of their Watson-Migdal FSI amplitude 
is at variance with that given in Ref. \cite{wat52} and by 
other authors \cite{shy99,shy98,dub86,mol96}.

In the second model used in Ref. \cite{sib00})(called as resonance
model in \cite{sib97}), the strangeness production proceeds via 
$\pi$-, $\eta$-, and $\rho$- exchange processes and the excitation 
of intermediate baryonic resonant states of $N^*$(1650), 
$N^*$(1710), $N^*$(1720), and $\Delta$(1920). In this case too,
with FSI effects included, they get the similar result for R. 
However, the excitation of the $N^*$(1650) baryonic
resonance has not been included in the calculations
of the cross sections for the $\Sigma^0K^+$ reaction in 
\cite{sib00}. In the near threshold region, the $\Lambda K^+$
reaction has been shown \cite{shy99,col97} to be 
dominated by this resonance. There is no $\it {a\,\, priori}$
reason to believe that it will not be the same for the $\Sigma^0K^+$
reaction.  

In this paper, we investigate the $\Lambda K^+$ and
$\Sigma^0 K^+$ reactions at 
near threshold as well as higher beam energies 
in the framework of an effective Lagrangian approach (ELA)
\cite{shy99,shy98,shy96}. In this model, the 
initial interaction between two incoming nucleons is 
modeled by an effective Lagrangian which is based on the exchange
of the $\pi$-, $\rho$-, $\omega$-, and $\sigma$- mesons. The coupling
constants at the nucleon-nucleon-meson vertices are determined by 
directly fitting the T-matrices of the nucleon-nucleon ($NN$) scattering
in the relevant energy region~\cite{sch94}. The ELA 
uses the pseudovector (PV) coupling for the nucleon-nucleon-pion
vertex (unlike the resonance model) and thus incorporates the low
energy theo\-rems of current algebra and the hypothesis of partially
conserved axial-vector current (PCAC). In contrast with the resonance
model, both the $\Lambda K^+$ and $\Sigma^0K^+$ reactions
proceed via excitation of the $N^*$(1650), $N^*$(1710), and $N^*$(1720)
intermediate baryonic resonance states. 
The interference terms between the amplitudes of various resonances
(which are ignored in \cite{sib00}) are retained. To describe the
near threshold data, the FSI effects in the final channel 
are included within the framework of the Watson-Migdal
theory~\cite{wat52,shy98}. ELA has been used earlier to describe 
rather successfully the $pp \to pp\pi^0$ and $pp \to pn\pi^+$
\cite{shy98,shy96} as well as $pp \to p\Lambda K^+$ \cite{shy99} 
reactions at both near threshold and higher beam energies.  

In the present form of the ELA 
the energy dependence of
the cross section due to FSI is separated from that of the primary 
production amplitude and
the total amplitude is written as,
\begin{eqnarray}
A_{fi} & = & M_{fi}(pp \rightarrow pYK^+) \cdot T_{ff},
\end{eqnarray} 
where $M_{fi}(pp \rightarrow pYK^+)$ is the primary associated hyperon
$YK^+$ production amplitude, while $T_{ff}$ describes the re-scattering 
among the final particles which goes to unity in the limit of no FSI. 
The latter is taken to be the coherent sum of the two-body on-mass-shell
elastic scattering amplitudes $t_i$ (with $i$ going from 1 to 3),
of the interacting particle pairs $j-k$ in the final
channel. This type of approach has been used earlier to 
describe the pion \cite{shy98,dub86,mei98}, $\eta$-meson\cite{mol96,dru97},
$\Lambda K^+$ \cite{shy99} and $\phi$-meson \cite{tit00} production
in $pp$ collisions.

An assumption inherent in Eq. (1) is that the reaction takes place over
a small region of space (which is fulfilled rather well in
near threshold reactions involving heavy mesons). Under this condition the
amplitudes $t_i$ can be expressed in terms of the inverse of the
Jost function $J_{\ell_i}(q_i)$ \cite{wat52,shy98}. Assuming  
the relative orbital angular momentum between  
pairs $j-k$ to be zero and using a (Coulomb) modified formula 
\cite{noy72} for the effective range expansion of the phase-shift
of the relevant pair, we can write \cite{wat52},
\begin{eqnarray}
t_i(q_i)  =  (J_0(q_i))^{-1} = \frac{(q_i^2+\alpha_i^2)r_{0i}^c/2}
                           {1/a_{i}^c+(r_{0i}^c/2)q_i^2-iq_i},
\end{eqnarray}
where $\alpha$ is defined as  
\begin{eqnarray}
\alpha & = & (1/r_{0i}^c)[1+(1+2r_{0i}^c/a_{i}^c)^{1/2}],
\end{eqnarray} 
with $a_i^c$ and $r^c_{0i}$ being the Coulomb modified \cite{shy99} 
effective range ($r_{0i}$) and scattering length ($a_{0i}$) 
parameters respectively and $q_i$ the relative momentum for the $j-k$ 
interacting pair. It is clear that for 
large $q_i$, the amplitude $t_i$ goes to unity. 
It should be noted that the form of the Jost function given in
\cite{sib00} does not lead to Eq. (2). Even though the square of
the absolute value of Eq. (2) agrees with that of the corresponding
function given in \cite{sib00}, the two forms lead to different results
if the FSI corrections in more than one final channel are considered. 

The validity of the factorization method (Eq. (1)) for applications to
the near threshold meson production in $pp$ collisions has recently been
investigated in Refs. \cite{han99,ged99,nis99,bar00}. 
It has been shown in Ref. \cite{ged99} that cross sections for the
$pp \to pp\pi^0$ reaction calculated using Eq. (1) 
are very similar to those obtained by treating $M_{fi}$ as an
effective operator acting on the nucleon wave functions 
calculated with realistic $NN$ interactions.
Furthermore, it is noted in
Ref. \cite{nis99} that for terms where $\pi^0$ production
proceeds via exchange of mesons between the colliding nucleons, the
results of the factorization approximation are quite similar to those
obtained by the DWBA calculations. It is
only for the direct (bremsstrahlung) terms (which are not included in
$M_{fi}$), that the results of the two calculations differ from each other
appreciably.

On the other hand, it is argued in Refs \cite{han99,bar00} that
although the energy dependence of the production process may be
described correctly by Eq. (1) (particularly for the production of heavier
mesons), its absolute magnitude could be 
uncertain because of the off-shell effects at the production vertices.
We have accounted for these effects by an off-shell
extrapolation of the on-shell FSI amplitude by multiplying it by
a monopole form factor with a cut-off parameter of 0.2 GeV, as suggested in
\cite{lag91,dru97}. In this method, both absolute
magnitude as well as shape of the FSI factor are affected by the 
off-shell corrections. In our calculations, the difference between the
off-shell and on-shell FSI factors is similar to that seen 
in Ref. \cite{bar00} for the case of Yamaguchi potential calculations of
the $\eta$-meson. The form factor approach for the off-shell effects
used here is based essentially on the Yamaguchi type
of separable $YN$ potential. It may be improved by using the off-shell
structure of some more realistic interaction. However, this will 
imply going beyond the factorization approach of Eq. (1),
which is beyond the scope of this paper.  
 
In our calculation of the FSI amplitudes, the values of the parameters
$a_0$ and $r_0$ for the $\Lambda$-$p$ and the $\Sigma$-$p$ systems were
taken from the $\tilde{A}$ model of the $YN$ interaction
of the J\"ulich-Bonn group \cite{reu94}. The values for these
parameters for the $\Lambda$-$p$ system were the same as 
those given in Ref. \cite{shy99}, while for the  
$\Sigma$-$p$ system, $a_0$ and $r_0$ were 2.28 fm and 4.96 fm for the
singlet state, and 0.76 fm and 2.50 fm for the triplet states respectively.
We have also considered the FSI interaction in the $K^+$-$Y$ channel,
which is possible only within the factorization approach
that has the additional advantage of making it 
possible to account for the FSI effects among all the three particles
in the outgoing channel. Since different two-body FSI amplitudes in the
final channel contribute coherently, the baryon-meson interactions, 
although weaker on their own, may still be influential through the 
interference terms. The values \cite{feu98,wal00}
of $a_0$ and $r_0$ were -0.065-$\it {i}$040 and
-15.930-$\it {i}$8.252 respectively for the $K^+$-$\Lambda$ system, and
-0.201-$\it {i}$0.131 and -1.757-$\it {i}$0.0835 respectively for 
the $K^+$-$\Sigma^0$ system. 
 
The amplitude $M_{fi}$ for the 
two reactions has been calculated in a way similar to that 
described in Ref. \cite{shy99} using the same set of
parameters.  However, we additionally require the coupling constants for the
$N^*\Sigma^0 K^+$ vertices  
in the calculation of the cross sections for 
the $\Sigma^0 K^+$ reaction. For $N^*$(1710) and $N^*$(1720)
resonances, these were determined from the corresponding branching
ratios (adopted from Ref. \cite{pdg98}) for their decay to the $\Sigma K$
channel. While choosing their values, 
we ensured that the sum of the branching ratios of all
the relevant channels does not exceed unity. The resulting coupling
constants are  given by $g^2_{N^*\Sigma^0K^+}/{4\pi}$ = 8.242 and 0.220 for
$N^*$(1710) and $N^*$(1720) resonances respectively with their signs
being negative and positive respectively.

However, such a procedure can not be used to determine $g_{N^*\Sigma^0K^+}$
for $N^*$(1650), as the on-shell decay of this resonance to the
$\Sigma K$ channel is inhibited. Instead, we tried to determine this 
coupling constant by fitting the available data on the 
$\pi^+p \to \Sigma^+ K^+$, $\pi^-p \to \Sigma^0 K^0$, and
$\pi^-p \to \Sigma^-K^+$ reactions in an effective Lagrangian  coupled
channels approach \cite{feu98,wal00}, where all the available data for the 
transitions from $\pi N$ to five meson-baryon final states, $\pi N$,
$\pi \pi N$, $\eta N$, $K\Lambda$, and $K\Sigma$ are simultaneously
analyzed for center of mass energies ranging from threshold to 2 GeV. In 
this analysis all the baryonic resonances with spin $\leq \frac{3}{2}$ 
up to excitation energies of 2 GeV are included as intermediate states. %
The best fit resulted in a value of 0.233 for the $N^*(1650)K\Sigma$ 
coupling, but due to very few data points available for the 
$\pi^- p\to K^+\Sigma^-$ channel, 
it may still be premature to attach much significance to this 
value. On the other hand, a value of 0.450 provides 
a very nice agreement with the data of the  $\Sigma^0 K^+$ 
reaction. Furthermore, since fitting with a fixed value of 0.450 to 
the available $\pi p \to K\Sigma$ data with the model mentioned above 
resulted in a comparable overall $\chi^2$ (although the former value 
provides a somewhat lower $\chi^2$ for $\pi^- p \to K^+\Sigma^-$, cf. 
Fig. 1), the latter value has been used in all the results shown in 
this paper. The shapes of the form factors and the values of the
cut-off parameters 
appearing therein were taken to be the same as those 
used in the case of $\Lambda K^+$ reaction \cite{shy99}.
 
In Fig. 2 we show the individual contributions of various nucleon
resonances to the total cross section of the
$\Sigma^0 K^+$ reaction near the production threshold as a function
of $\epsilon$. We see that, as in the case of the
$\Lambda K^+$ reaction, the
cross section for this reaction too is dominated by the $N^*$(1650)
resonance excitation. Thus, at the near threshold
energies, both these reactions proceed preferentially via excitation
of this resonance. Looking only at the values of 
coupling constants, one might expect the dominance of
$N^*$(1710) resonance for both the reactions even at these energies. 
This is particularly so for the $\Sigma^0K^+$
reaction, where the threshold energy is very close to the excitation
energy of this resonance.  However, in the near threshold region the 
relative dominance of various resonances is determined by the
dynamics of the reaction where the difference of about 60 MeV in
excitation energies of $N^*$(1650) and $N^*$(1710) resonances plays a
crucial role. Yet, some differences in the relative contributions
of $N^*$(1710) resonance in the two reactions at these 
energies are noteworthy. For 
$\Sigma^0K^+$ reaction the contribution of this resonance
is about a factor of 3-4 larger as compared to that in the case
of $\Lambda K^+$ reaction. This is the reason for the interference
effects among the resonances being relatively larger in Fig. 2 as 
compared to that in the $\Lambda K^+$ case \cite{shy99}. 
It may be remarked here that
in both cases one-pion exchange between the incident protons gives maximal 
contribution to the cross sections as compared to $\rho$-, $\omega$-, and 
$\sigma$-meson exchanges.

The total cross sections for the $\Lambda K^+$ and $\Sigma^0 K^+$ reactions
as a function of $\epsilon$ are shown in Fig. 3. The calculations are
the coherent sum of all resonances and meson exchange processes
as described earlier. The $\Lambda K^+$ results are the same as 
those shown in \cite{shy99}. For the $\Sigma^0K^+$ reaction,
there is a reasonable agreement between theory and the data except
for very small values of $\epsilon$ where our calculations
underpredict the experimental cross sections by a factor of about 1.5.
Keeping in mind the fact that all parameters of the model, except
for those of $N^*Yp$ vertices and the FSI, were the same in the two
calculations and that no parameter was freely varied, this agreement is
quite satisfactory. It should be noted that unlike Ref. \cite{gas00},
we do not require to introduce arbitrary normalization constants 
to get the agreement between calculations and the data.
 
In Fig. 4, we compare our calculations with the data
for the ratio R as a function of $\epsilon$. We have shown here the
results for excess energies up to 1 GeV, where the first high energy data
is available. It is clear that our calculations are able to describe the
strong fall-off of R between low and high energies even though they
somewhat overestimate the effects at the lowest points.  
It is interesting to note that at the near threshold energies,
calculations done without FSI effects can already explain the data
up to 40-50$\%$. Therefore, all of the observed value of
R at these beam energies can not be accounted for by the FSI alone, which
is in agreement with the observation made in \cite {gas00}. It should
again be emphasized that without considering the contributions of the
$N^*$(1650) resonance for the $\Sigma^0K^+$ reactions the
calculated ratio would be at least an order of magnitude larger.
Therefore, these data are indeed sensitive to the details of the
reaction mechanism. At higher beam energies ($\epsilon$ $>$ 300 MeV),
values of R obtained with and without FSI effects are almost identical.
In this region the reaction mechanism is different; here the $N^*$(1710)
resonance makes the dominant contribution \cite{shy99} and FSI related
effects are unimportant. This is the most likely cause for the
difference in the values of R in the two regions.

In summary, we have studied the $pp \to p\Lambda K^+$ and 
$pp \to p\Sigma^0K^+$ reactions within an effective Lagrangian model.
Most of the parameters of the model are fixed by fitting the
$NN$ T-matrix, which restricts the freedom of varying them freely in order 
to fit the data. The reactions proceed via the excitation of the 
$N^*$(1650), $N^*$(1710), and $N^*$(1720) intermediate baryonic resonant
states. An important result of our study is that in the 
near threshold region both these reactions proceed predominantly via
excitation of the $N^*$(1650) intermediate baryonic resonant state. To
the extent that the final state interaction effects in the exit channel 
can be accounted for by the Watson-Migdal theory, our model is able to  
explain the experimentally observed large
ratio of the total cross sections of the two reactions in the near threshold
region. It can also explain the relatively smaller value of this ratio 
at higher beam energies where the reactions are dominated
by the $N^*$(1710) resonance and the FSI related effects are negligible.
   
One of the authors (RS) would like to thank Pawel Danielewicz for
his kind hospitality in the theory group of the National Superconducting
Cyclotron Laboratory of the Michigan State University where a part of 
this work was done.

\newpage
\begin{center} Figure Captions \end{center}
\begin{itemize}
\item [Fig. 1]
The total cross section for the
$\pi^-p \to \Sigma^0K^0$ (upper part) and $\pi^-p \to \Sigma^-K^+$ (lower 
part) reactions as a function of invariant mass s. The solid and dashed 
lines show the results of a coupled channels K-matrix calculation 
~\protect\cite{feu98} with the values of the coupling constant for
$N^*$(1650)$\Sigma K$ vertex of 0.233 and 0.450 respectively.

\item [Fig. 2]
Contributions of $N^*$(1650) (dotted line), $N^*$(1710)
(dashed line) and $N^*$(1720) (dashed-dotted line) baryonic resonances
to the total cross section for the 
$pp \to p\Sigma^0K^+$ reaction as a function of the excess energy. 
Their coherent sum is shown by the solid line. 

\item [Fig. 3]
Comparison of the calculated and the experimental total
cross section for the $pp \to p\Lambda K^+$ (solid line and
solid squares) and $pp \to p\Sigma^0K^+$ (dashed line and solid circles)
as a function of the excess energy. The experimental data are from
Refs. \cite{bal98} (solid squares) and \cite{sew99} (solid circles).

\item [Fig. 4]
Ratio of the total cross sections for 
$pp \to p\Lambda K^+$ and $pp \to p\Sigma^0 K^+$ reaction as a function
of the excess energy. The solid and dashed lines show the results of our
calculations with and without FSI effects respectively. The data are from
\cite{sew99,lan88}. 
\end{itemize}

\begin{figure}
\begin{center}
\mbox{\epsfig{file=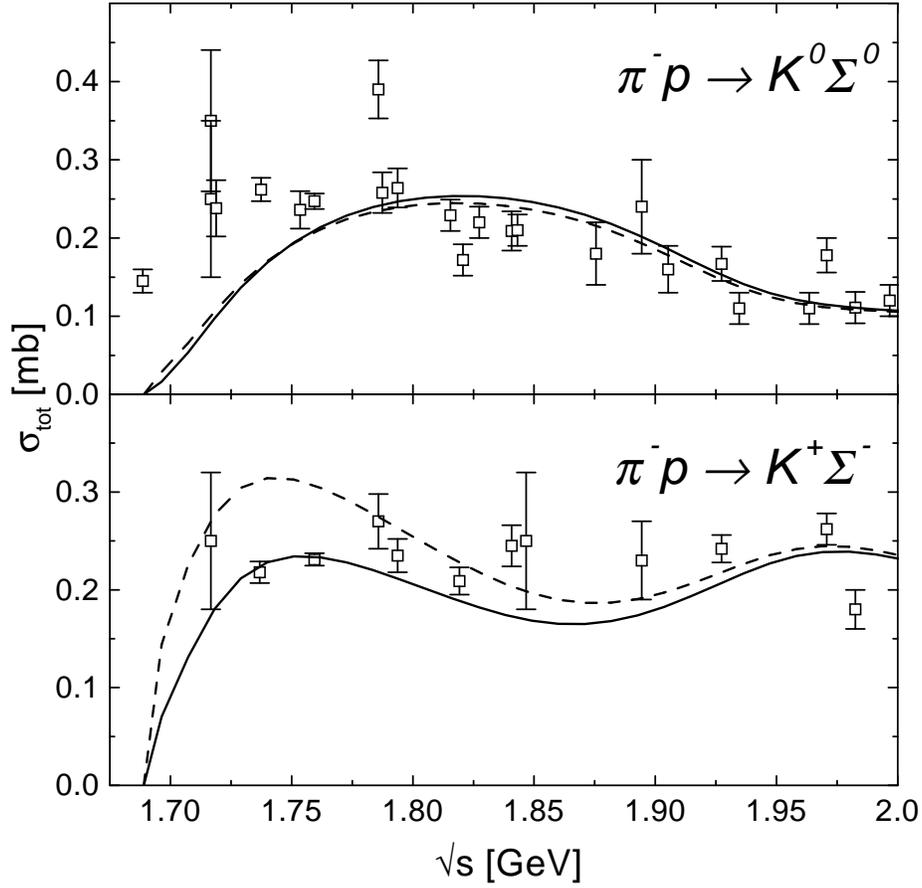,height=12.0cm}}
\end{center}
\caption {
The total cross section for the
$\pi^-p \to \Sigma^0K^0$ (upper part) and $\pi^-p \to \Sigma^-K^+$ (lower 
part) reactions as a function of invariant mass s. The solid and dashed 
lines show the results of a coupled channels K-matrix calculation 
~\protect\cite{feu98} with the values of the coupling constant for
$N^*$(1650)$\Sigma K$ vertex of 0.233 and 0.450 respectively.
 } 
\label{fig:figa}
\end{figure}

\begin{figure}
\begin{center}
\mbox{\epsfig{file=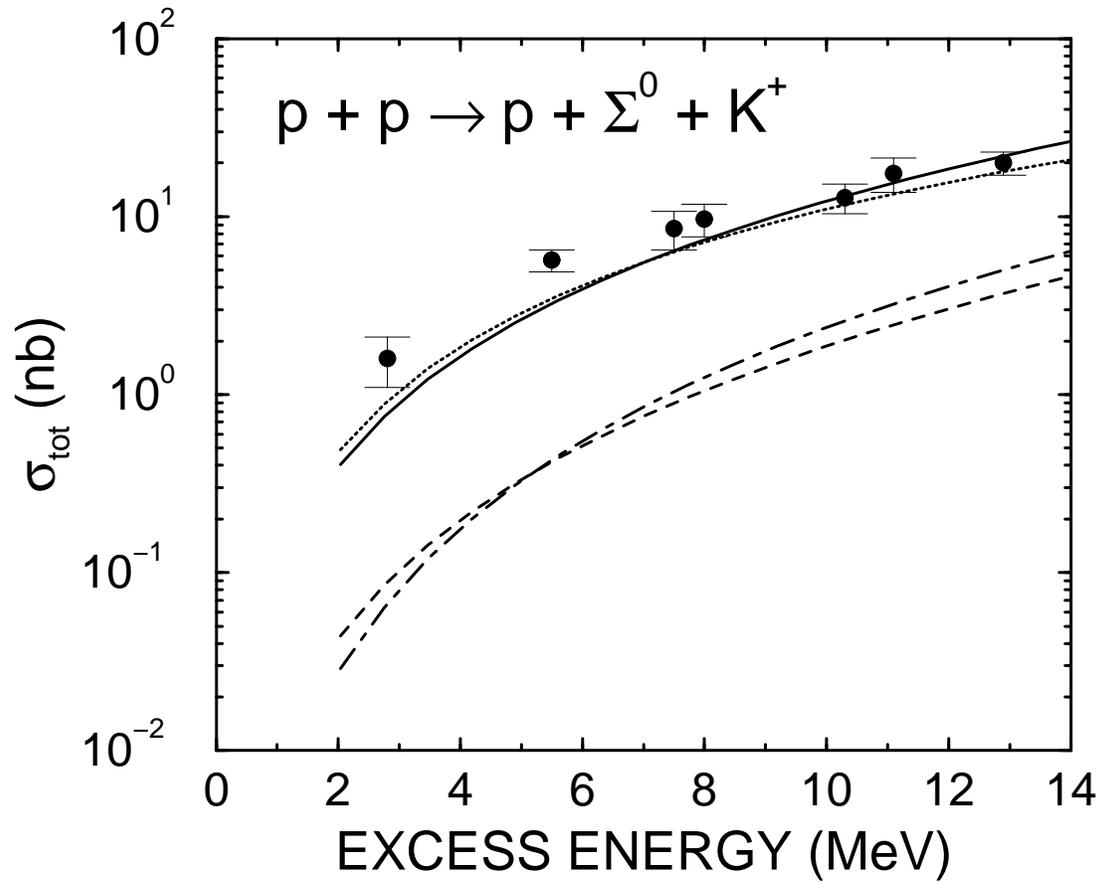,height=12.0cm}}
\end{center}
\caption {
Contributions of $N^*$(1650) (dotted line), $N^*$(1710)
(dashed line) and $N^*$(1720) (dashed-dotted line) baryonic resonances
to the total cross section for the 
$pp \rightarrow p\Sigma^0K^+$ reaction as a function of the excess energy. 
Their coherent sum is shown by the solid line. 
}
\label{fig:figb}
\end{figure}
\begin{figure}
\begin{center}
\mbox{\epsfig{file=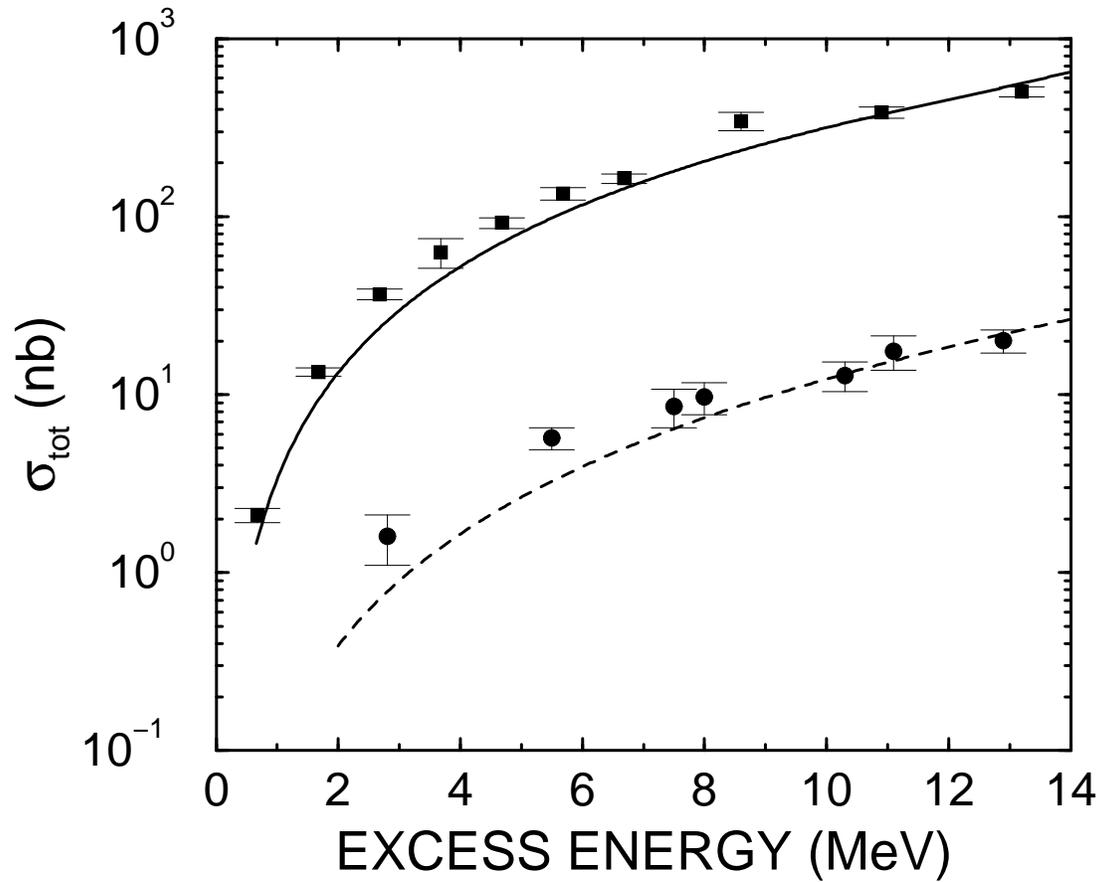,height=12.0cm}}
\end{center}
\vskip .3in
\caption{
Comparison of the calculated and the experimental total
cross section for the $pp \to p\Lambda K^+$ (solid line and
solid squares) and $pp \to p\Sigma^0K^+$ (dashed line and solid circles)
as a function of the excess energy. The experimental data are from
Refs. ~\protect\cite{bal98} (solid squares) and ~\protect\cite{sew99}
(solid circles).
}
\label{fig:figc}
\end{figure}

\begin{figure}
\begin{center}
\mbox{\epsfig{file=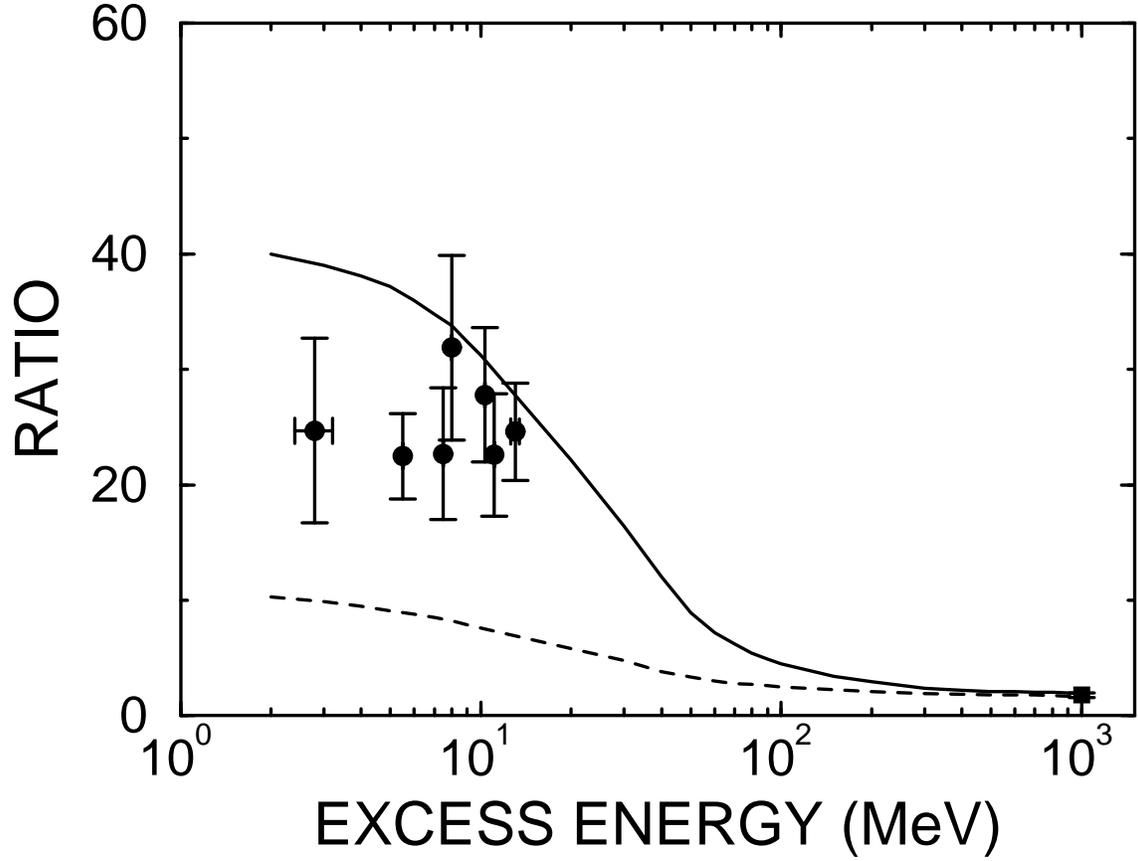,height=12.0cm}}
\end{center}
\vskip .3in
\caption {
Ratio of the total cross sections for 
$pp \to p\Lambda K^+$ and $pp \to p\Sigma^0 K^+$ reaction as a function
of the excess energy. The solid and dashed lines show the results of our
calculations with and without FSI effects respectively. The data are from
~\protect\cite{sew99,lan88}.
} 
\label{fig:figd}
\end{figure}
\end{document}